\documentclass[11pt]{article}
\usepackage{appb,epsfig}

\begin{document}

\title{The $\pi^-p\rightarrow\gamma\gamma n$ Reaction and Pion Compton Scattering \thanks{Presented at MESON'98} }
\author{ Piotr A. \.Zo{\l}nierczuk \\ on behalf of the RMC collaboration\address{Dept. of Physics and Astronomy, University of Kentucky, Lexington, USA}}

\maketitle

\begin{abstract}
We propose  to measure 
the two--photon capture mode of pionic hydrogen 
using the RMC \cite{Wr92} pair spectrometer at the TRIUMF cyclotron.
Currently, only an experimental upper limit of 
$B.R.\leq 5.5\times10^{-4}$ is available
for the $\gamma\gamma$ capture mode ~\cite{Va69}.
Our new data will critically test the theoretical calculations 
and the intriguing predicted dominance of the
$\pi\pi\rightarrow\gamma\gamma$ annihilation graph.
More speculatively, since crossing symmetry relates
$\pi\pi\rightarrow\gamma\gamma$ to $\gamma\pi\rightarrow\gamma\pi$, 
this threshold reaction may be a novel probe of the
pion's electric polarizability.

We show results of our 1997 engineering run and Monte Carlo studies 
for the two--photon capture mode of pionic hydrogen and carbon.
We also present our preliminary upper limit for the 
$B.R.[\pi^-p\rightarrow\gamma\gamma n] \leq 2.8\times10^{-4}$.
\end{abstract}

\section{Current status of \boldmath{$\pi^-p\rightarrow\gamma\gamma n$}}
\label{sec:motivation}

The $\pi^0n$, $\gamma n$ and $e^+e^-n$ branching ratios of 
the pionic hydrogen atom are accurately measured 
(see Tab.\ref{table:capture modes}). 
For the $\gamma\gamma n$ mode, however, there is only 
an upper limit of the branching ratio, $B.R. \leq 5.5 \times 10 ^{-4}$, 
obtained in a JINR spark chamber experiment~\cite{Va69}. 
\begin{table}[htbp]
  \begin{center}
    \caption{Capture modes and branching ratios of the pionic hydrogen atom.}
    \begin{tabular}{lll}
      \hline
      Capture Mode & Branching Ratio & Reference \\
      \hline
      $\pi^-p \rightarrow \pi^o n$         & $0.607  \pm 0.004 $    & \cite{Sp77} \\
      $\pi^-p \rightarrow \gamma n$        & $0.386  \pm 0.002 $    & \cite{Sp77} \\
      $\pi^-p \rightarrow e^+e^- n$        & $0.0069 \pm 0.0003$    & \cite{Sa61} \\
      $\pi^-p \rightarrow \gamma \gamma n$ & $\leq5.5\times10^{-4}$ & \cite{Va69} \\
      \hline
    \end{tabular}
    \label{table:capture modes}
  \end{center}
\end{table}

Although the hydrogen ($\pi^-,\gamma\gamma$) 
reaction has not been observed, the nuclear  
($\pi^-,\gamma\gamma$) reaction has been measured at PSI and TRIUMF \cite{De79}.
The branching ratios obtained from the two carbon ($\pi^-$,$\gamma\gamma$) 
experiments are in reasonable agreement,
Deutsch {\em et.al} obtained $(1.4 \pm 0.2) \times 10 ^{-5}$ 
and  Mazzucato {\em et.al} $(1.2 \pm 0.2) \times 10 ^{-5}$. 

Several authors have made tree--level calculations of the $\gamma\gamma n$ capture mode
of pionic hydrogen.
For the $\gamma\gamma n$ branching ratio Joseph obtained 5.1$\times$10$^{-5}$,
Lapidus and Musakhanov 4.0$\times$10$^{-5}$, 
and Beder 5.1$\times$10$^{-5}$~\cite{Be79}.
These values are roughly ten times smaller than the experimental upper limit 
from the JINR experiment.
Beder also pointed out the importance of the pion annihilation diagram,
especially at small photon opening angles (see Table 3 and Fig. 4 of Beder paper~\cite{Be79}). 

\section{Relation to Pion Compton scattering}
\label{sec:compton}

\begin{figure}[htbp]
    \leavevmode
    \begin{center}
    \begin{tabular}{ccc}
      \epsfig{width=3cm,height=2.25cm,figure=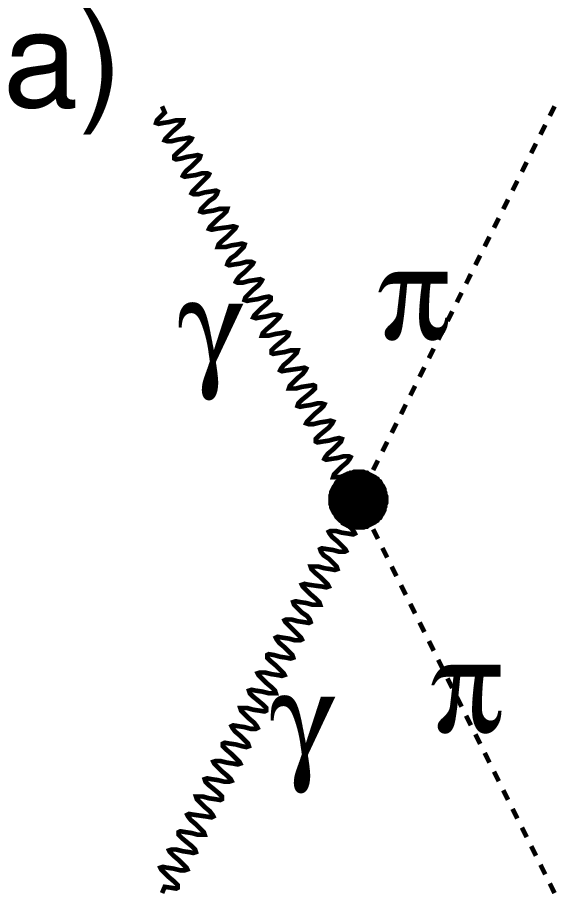} & 
      \epsfig{width=3cm,height=2.25cm,figure=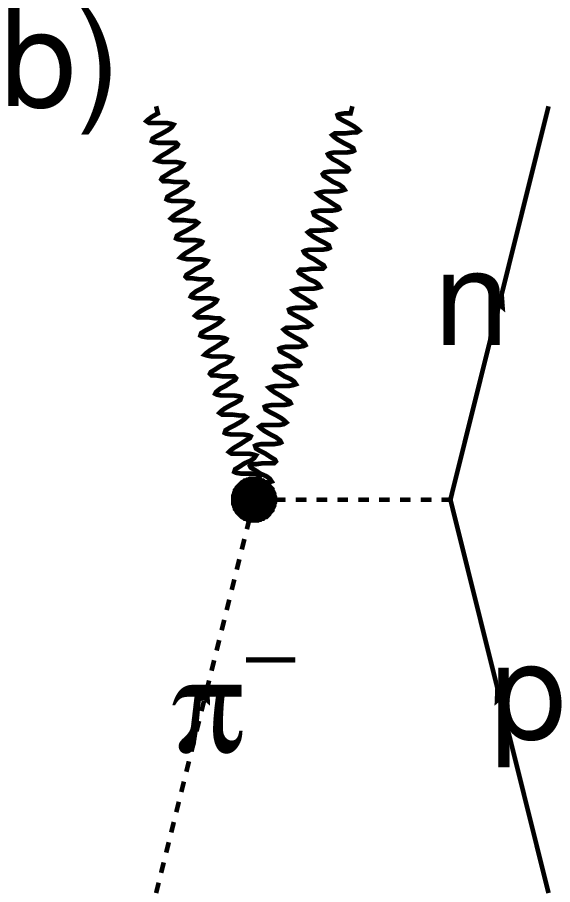} & 
      \epsfig{width=3cm,height=2.25cm,figure=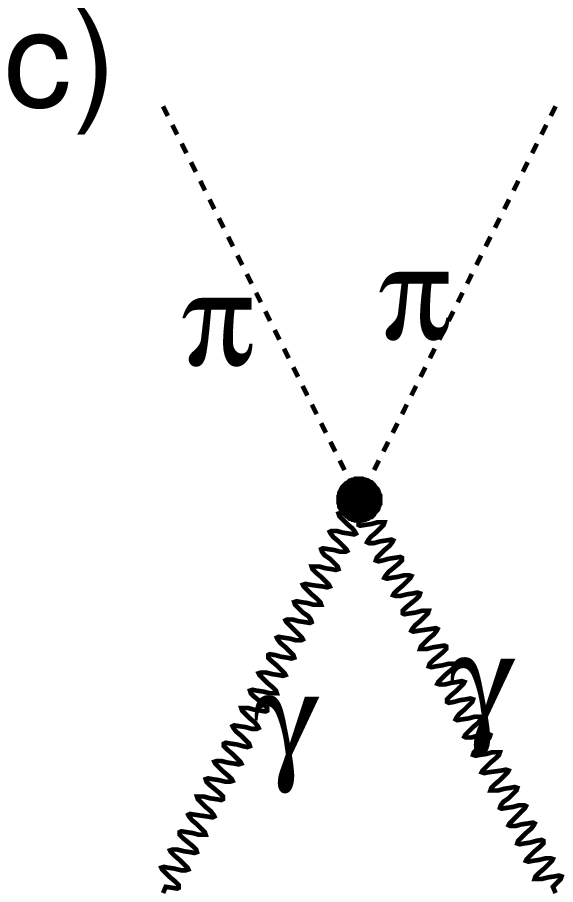} \\
      $ t        \le 0 $            & 
      $ 0        \le t \le m_\pi^2$ &
      $ 4m_\pi^2 \le t $            \\
    \end{tabular}
    \caption{a) Pion Compton scattering, 
      b) $\pi^-p\rightarrow\gamma\gamma n$ annihilation graph 
      c) $\gamma\gamma$ $\rightarrow \pi\pi$}
    \label{fig:pioncompton}
    \end{center}
\end{figure}

An intriguing feature of the $\gamma\gamma n$ capture mode 
is the predicted dominance of the $\pi\pi\rightarrow\gamma\gamma$ 
annihilation diagram (Fig.\ref{fig:pioncompton}b). 
One can view this Feynman diagram as the annihilation of a real 
pion with a virtual pion
$\pi^-\pi^+\rightarrow\gamma\gamma$ or, via crossing symmetry,
as the transition of a real pion to a virtual pion via Compton scattering 
$\gamma\pi\rightarrow\gamma\pi$ (Fig.\ref{fig:pioncompton}a).
The four--momentum transfer--squared in 
the $\pi^-p\rightarrow\gamma\gamma n$ reaction
is $0<t<+m^2_{\pi}$, compared with
$t\leq 0$ for the on--shell $\gamma\pi$ $\rightarrow$ $\gamma\pi$ (Fig.\ref{fig:pioncompton}a) 
process and $t\geq 4m^2_{\pi}$ for the on--shell 
$\gamma\gamma$ $\rightarrow \pi\pi$ (Fig.\ref{fig:pioncompton}c) process. 

Pion Compton scattering is a probe of pion's polarizability 
$\alpha_E^{\pi^\pm}$ (see e.g.\cite{Ho90}).
A summary of the current determinations  of the pion's polarizability
is shown in Tab.~\ref{tab:pionpol}.
The extracted pion polarizabilities have 
large uncertainties and are (except MARK II)
substantially larger than the `$\chi PT$ prediction \cite{Ho90},
$\alpha_E^{\pi^\pm} = (2.7 \pm 0.4)\times 10^{-4}\;fm^3$,
which is based on a relationship between
radiative pion decay $\pi\rightarrow e\nu\gamma$
and pion Compton scattering $\gamma\pi$ $\rightarrow$ $\gamma\pi$.
This discrepancy between theory and experiment calls for more experimental attention,
thus our proposal to measure two--photon capture mode of the pionic hydrogen
appears timely.
\begin{table}[htbp]
  \begin{center}
    \leavevmode
    \caption{Experimental determinations of pion electric polarizability $\alpha_E^{\pi^\pm}$}
    \begin{tabular}{lll}
      \hline
      Experiment & $\alpha_E^{\pi^\pm}\; (\times 10^{-4}\;fm^3)$ & Reference \\
       \hline
      $\pi A \rightarrow \gamma \pi A $    & $6.8 \pm 1.4 \pm 1.2 $  & Serpukhov \cite{An84}\\
      $\pi p \rightarrow \gamma \pi p $    & $20 \pm 12 $            & Lebedev   \cite{Ai86}\\
      $\gamma \gamma \rightarrow \pi \pi $ & $19.1 \pm 4.9 \pm 5.6 $ & PLUTO     \cite{Ba92}\\
      $\gamma \gamma \rightarrow \pi \pi $ & $2.2 \pm 1.6 $          & MARK II   \cite{Ba92}\\
      \hline
    \end{tabular}
    \label{tab:pionpol}
  \end{center}
\end{table}

\section{TRIUMF \boldmath{$\pi^-p\rightarrow\gamma\gamma n$}  experiment}
\label{sec:triexp}

Using the RMC pair spectrometer~\cite{Wr92} we propose to measure the 
$\pi^-p\rightarrow\gamma\gamma n$ branching ratio and its 
photon opening angle and energy partition dependence.
\begin{figure}[ht]
  \begin{center}
    \epsfig{figure=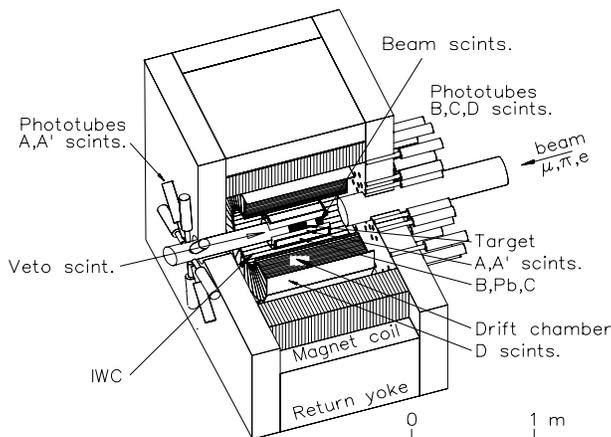,angle=0,width=8.0cm} 
    \caption{Layout of the RMC photon pair spectrometer
      showing the liquid hydrogen target, Pb converter,
      cylindrical wire and drift chambers, 
      trigger scintillators and solenoidal magnet.}
  \end{center}
  \label{fig:rmcspectrometer}
\end{figure}

We have performed an extensive experimental and Monte Carlo
studies of both pionic hydrogen and pionic carbon
two--photon capture modes that included trigger tuning, 
acceptance optimization and background suppression~\cite{Go98}.
These studies have proved the feasibility of the proposed experiment.
\begin{figure}[htbp]
  \begin{center}
    \leavevmode
    \epsfig{width=10cm,height=5cm,figure=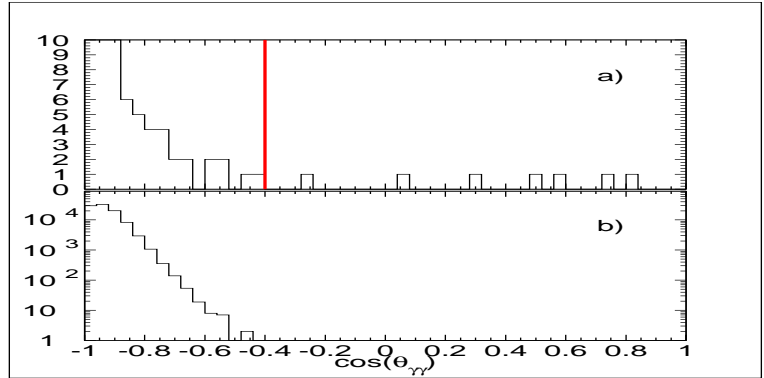}    
    \caption{Two--photon opening angle spectra: a) for pionic carbon and b) for pionic hydrogen}
    \label{fig:signal}
  \end{center}
\end{figure}

Fig.~\ref{fig:signal} presents our preliminary results taken  
during our 1997 engineering run. Both spectra show two--photon opening angle distribution,
the upper one for pionic carbon, the lower one for pionic hydrgen.
The events with large opening angles ($\cos\theta{\gamma\gamma}\le -0.4$) are 
photons from $\pi^0$ decay originating from the $\pi^-p\rightarrow\pi^0 n$ 
charge exchange reaction.

We have found 7 two--photon 
events from pionic carbon with small opening angles ($\cos\theta{\gamma\gamma}\ge -0.4 $,
see Fig.~\ref{fig:signal}a) 
which yields $B.R.[^{12}C(\pi^-,\gamma\gamma)]=(0.8 \pm 0.3\,\mbox{(stat.)}) \times 10^{-5}$
in reasonable agreement with \cite{De79}.
We also have searched for small opening angle two--photon from events pionic hydrogen  
and so far found none (see Fig.~\ref{fig:signal}b), 
which results  in an improved upper limit of 
$B.R.[\pi^-p\rightarrow\gamma\gamma n] \leq 2.8 \times 10^{-4}$.

\bibliographystyle{unsrt}

\end{document}